\documentstyle[sprocl]{article}
\include{epsf}
\def\Journal#1#2#3#4{{#1} {\bf #2}, #3 (#4)} 
\def\NCA{\em Nuovo Cimento} 
 
\def\PLB{{\em Phys. Lett.}  B} 
 
\def\PRD{{\em Phys. Rev.} D} 
 
\begin{document}
\hfill Preprint Numbers:\parbox[t]{45mm}{ADP-96-45/T239\\
                         FSU-SCRI-96-135\\hep-ph/9612360}\\
\title{Subtractive Renormalization of Strong-Coupling Quenched QED
in Four Dimensions}

\author{ Frederick T.\ Hawes}

\address{Department of Physics and SCRI, Florida State University,\\
       Tallahassee, Florida 32306-3016 USA }

\author{Tom Sizer, Anthony G.\ Williams}

\address{Department of Physics and Mathematical Physics,\\
       University of Adelaide, South Aust. 5005, Australia }

%
\maketitle\abstracts{
  We study renormalized quenched strong-coupling QED in four dimensions
  in arbitrary covariant gauge, in the Dyson-Schwinger equation
  formalism.  Above the chiral critical coupling, we show that
  there is no finite chiral limit.  This behaviour is found to be
  independent of the detailed choice of proper vertex,
  provided that the vertex is consistent with
  the Ward-Takahashi identity and multiplicative renormalizability.
  The finite solutions previously reported 
  lie in an unphysical regime of the theory with multiple
  solutions and oscillating mass functions.
  This study is consistent with the assertion that strong coupling
  QED$_4$ does not have a continuum limit in the conventional sense.
}

\section{Introduction}
\label{sec_intro}

The abelian nature of quantum electrodynamics (QED)
in many ways makes it a much simpler system to study
than a nonabelian theory such as quantum chromodynamics (QCD).
For this reason it has been the subject of many nonperturbative
studies, which have as their long-term goal a detailed understanding
of nonperturbative QCD.~\cite{TheReview,MiranskReview,FGMS}
On the other hand, strong-coupling QED$_4$ is widely anticipated
to behave unconventionally in the continuum limit
and for this reason is a theory of considerable interest
in its own right.

In previous work \cite{qed4_hw,qed4_hwr} we introduced a
numerical renormalization procedure for the Dyson-Schwinger equations,
and applied it to QED$_4$ with a quenched photon propagator,
under momentum cutoff regularization.
The central result of those works was to demonstrate that
the numerical renormalization procedure works extremely well
and allows the continuum limit (cutoff $\Lambda\to\infty$) to be taken
numerically, while giving rise to stable finite solutions
for the renormalized fermion propagator.

Here we describe studies of the chiral limit
in renormalized quenched strong-coupling QED$_4$,\cite{qed4_HSW}
using a photon-fermion vertex
that satisfies the Ward-Takahashi Identity (WTI)
and makes the fermion DSE multiplicatively renormalizable.
We find that for couplings above the chiral critical coupling,
keeping the bare mass $m_0(\Lambda) \equiv 0$
as the cutoff is relaxed results in either
a dynamical mass function which is identically zero,
or a dynamical mass function which diverges with the cutoff.  
The previously reported finite solutions showed damped oscillations
in the dynamical mass functions at large $p^2$,
which suggested that they were unphysical.
Further, we show here that for a given supercritical coupling
and the same bare mass $m_0(\Lambda)$,
it is possible to have multiple solutions
corresponding to different renormalized masses $m(\mu)$.
We conclude that above the chiral phase transition,
quenched strong-coupling QED in four dimensions
does not have a chiral limit in the conventional sense.

\section{Formalism}
\label{sec_formalism}


We write the fermion propagator as
\begin{equation}                        \label{fermprop_formal}
  S(p) = \frac{Z(p^2)}{\not\!p - M(p^2)}
       = \frac{1}{A(p^2) \not\!p - B(p^2)}
\end{equation}
where we refer to $A(p^2)\equiv 1/Z(p^2)$
as the finite momentum-dependent fermion renormalization
and where $M(p^2)\equiv B(p^2)/A(p^2)$ is the fermion mass function.
Dynamical chiral symmetry breaking (DCSB) occurs when the
fermion propagator develops a nonzero scalar
self-energy in the absence of an explicit chiral symmetry breaking
(ECSB) fermion mass.
We will refer to coupling constants strong enough
to induce DCSB as supercritical and those weaker as subcritical.

The DSE for the renormalized fermion propagator,
in an arbitrary covariant gauge, is
\begin{equation} \label{fermDSE_eq}
  S^{-1}(p) =  Z_2(\mu,\Lambda)[\not\!p - m_0(\Lambda)]
    -i Z_1(\mu,\Lambda) e^2 \int^{\Lambda} \frac{d^4k}{(2\pi)^4}
	  \gamma^{\mu} S(k) \Gamma^{\nu}(k,p) D_{\mu \nu}(q)\:;
\end{equation}
here $q=k-p$ is the photon momentum, $\mu$ is the renormalization
point, and $\Lambda$ is a regularizing parameter (taken here to be an
ultraviolet momentum cutoff), and $m_0(\Lambda)$ is the
regularization-parameter dependent bare mass.
We work in the quenched approximation,
i.e., the photon propagator is bare,
so that for the coupling strength and gauge parameter we have
$\alpha\equiv e^2/4\pi = \alpha_0\equiv e_0^2/4\pi$ and $\xi=\xi_0$.


Among other general requirements \cite{TheReview,Craig_1}, the vertex
{\it Ansatz\/} is constrained by the Ward-Takahashi Identity (WTI),
which is necessary for gauge invariance and guarantees the equality
of the propagator and vertex renormalization constants,
$Z_2 \equiv Z_1$.
Ball and Chiu \cite{BC} have described the most general
fermion-photon vertex that satisfies the WTI;
it consists of a longitudinally-constrained (i.e., ``Ball-Chiu'') part
$\Gamma^\mu_{\rm BC}$, which is a minimal solution of the WTI
with no artificial kinematic singularities,
and a set of eight vectors $T_i^\mu(k,p)$,
which span the transverse subspace and are also kinematically
regular as $k \to p$ (improved versions of the $T_i^\mu$
can be found in Ref.~\cite{Kiz_et_al}).
A general vertex is then written as
  $\Gamma^\mu(k,p) = \Gamma_{BC}^\mu(k,p)
    + \sum_{i=1}^{8} \tau_i(k^2,p^2,q^2) T_i^\mu(k,p)\:$,
where the $\tau_i$ are functions that must be chosen to ensure
the correct discrete symmetry properties, perturbative
limits, etc.
For our numerical examples we have used the Curtis-Pennington
vertex \cite{CPItoIV}, however our results and conclusions
regarding the chiral limit will not depend on the specific
vertex {\it Ansatz} as long as it respects multiplicative
renormalizability.


The renormalization procedure is entirely analogous to
one-loop perturbative renormalization:
One first determines a finite, {\it regularized\/} self-energy,
$\Sigma'(\mu,\Lambda; p)$,
depending on both the regularization parameter $\Lambda$
and the renormalization point $\mu$.
The DSE for the renormalized fermion propagator can be written
in terms of $\Sigma'(\mu,\Lambda; p)$
as
\begin{eqnarray} \label{ren_inv_S}
  {S}^{-1}(p) & = & Z_2(\mu,\Lambda) [\not\!p - m_0(\Lambda)]
    - \Sigma'(\mu,\Lambda; p) \nonumber\\
    & = & \not\!p - m(\mu) - \widetilde{\Sigma}(\mu;p)
      = A(p^2)\not\!p -B(p^2)\:,
\end{eqnarray}
where
$\widetilde{\Sigma}(\mu;p)$ denotes the {\it renormalized} self-energy.
The self-energies are decomposed into Dirac and scalar parts,
  $\Sigma'(\mu,\Lambda; p) = \Sigma'_d(\mu,\Lambda; p^2) \not\!p
		     + \Sigma'_s(\mu,\Lambda; p^2)$\/,
(and similarly for $\widetilde{\Sigma}(\mu,p)$).
By imposing the renormalization boundary condition,
\begin{equation}
  \left. {S}^{-1}(p) \right|_{p^2 = \mu^2} = \not\!p - m(\mu)\:,
\label{ren_point_BC}
\end{equation}
one finds relations
for the scalar and Dirac self-energies, propagator renormalization constant
$Z_2(\mu,\Lambda)$, and bare mass.
A mass renormalization constant can also be defined as
  $Z_m(\mu,\Lambda) = m_0(\Lambda)/m(\mu)$,
i.e., as the ratio of the bare to renormalized mass.


We approach the chiral limit by setting the bare mass to zero
and removing the regularization, i.e., maintaining $m_0(\Lambda)=0$
while taking the limit $\Lambda\to\infty$;
this is the same definition used in nonperturbative studies
of QCD\cite{TheReview,MiranskReview,FGMS}.
Obviously, any limiting procedure where we take $m_0(\Lambda)\to 0$
sufficiently rapidly as $\Lambda\to \infty$ will also lead
to the chiral limit.


The behavior of the solutions under a renormalization point
transformation is governed by a ``renormalization group flow'':
under any transformation $\mu \to \mu'$, we must have
{\em for all } $p^2$
\begin{eqnarray}
  M(\mu';p^2)&=&M(\mu;p^2)\equiv M(p^2) \;, \nonumber\\
  \frac{A(\mu';p^2)}{A(\mu;p^2)}&=&
     \frac{Z_2(\mu',\Lambda)}{Z_2(\mu,\Lambda)}
     =A(\mu';\mu^2)=\frac{1}{A(\mu;\mu'^2)} \;,
\label{ren_pt_transf}
\end{eqnarray}
so that, for the fermion propagator,
  $S(\mu';p)/S(\mu;p) = Z_2(\mu,\Lambda)/Z_2(\mu',\Lambda)$
in the usual way.  We have verified Eq.~(\ref{ren_pt_transf})
to within 1 part in $10^4$ for our calculations.

The DSE is solved in Euclidean momentum-space,
after separation into Dirac-odd and -even parts
describing the finite renormalization $A(p^2)$
and the scalar self-energy $B(p^2)$.
We use a ``gauge-covariance-improved'' momentum cutoff scheme,
which is described in detail in Appendix A of Ref.~\cite{qed4_hwr}
Choosing the renormalized mass, $m(\mu)$, and then solving
for the bare mass, $m_0(\Lambda)$, leads to well-behaved
finite solutions for $A(p^2)$ and $M(p^2)$, which do not vary
as we take the continuum limit ($\Lambda\to\infty$)
\cite{qed4_hw,qed4_hwr}.

\section{Results and Conclusions}
\label{sec_results}

\begin{figure}[htb]
  \setlength{\epsfysize}{4.5cm}
      \epsffile{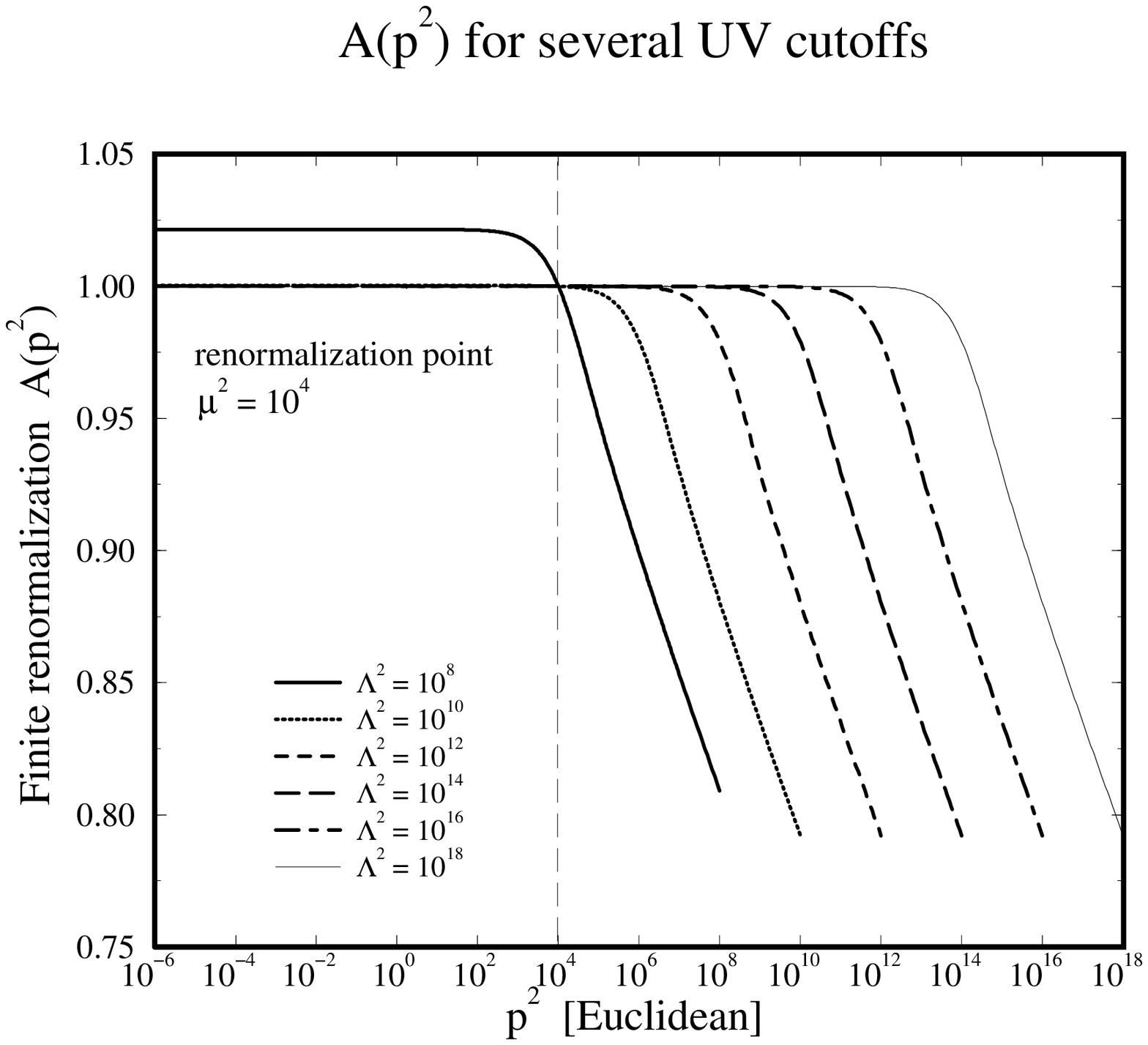}
  \vskip -5.0cm
  \null
  \hskip  5.5cm
  \setlength{\epsfysize}{4.5cm}
      \epsffile{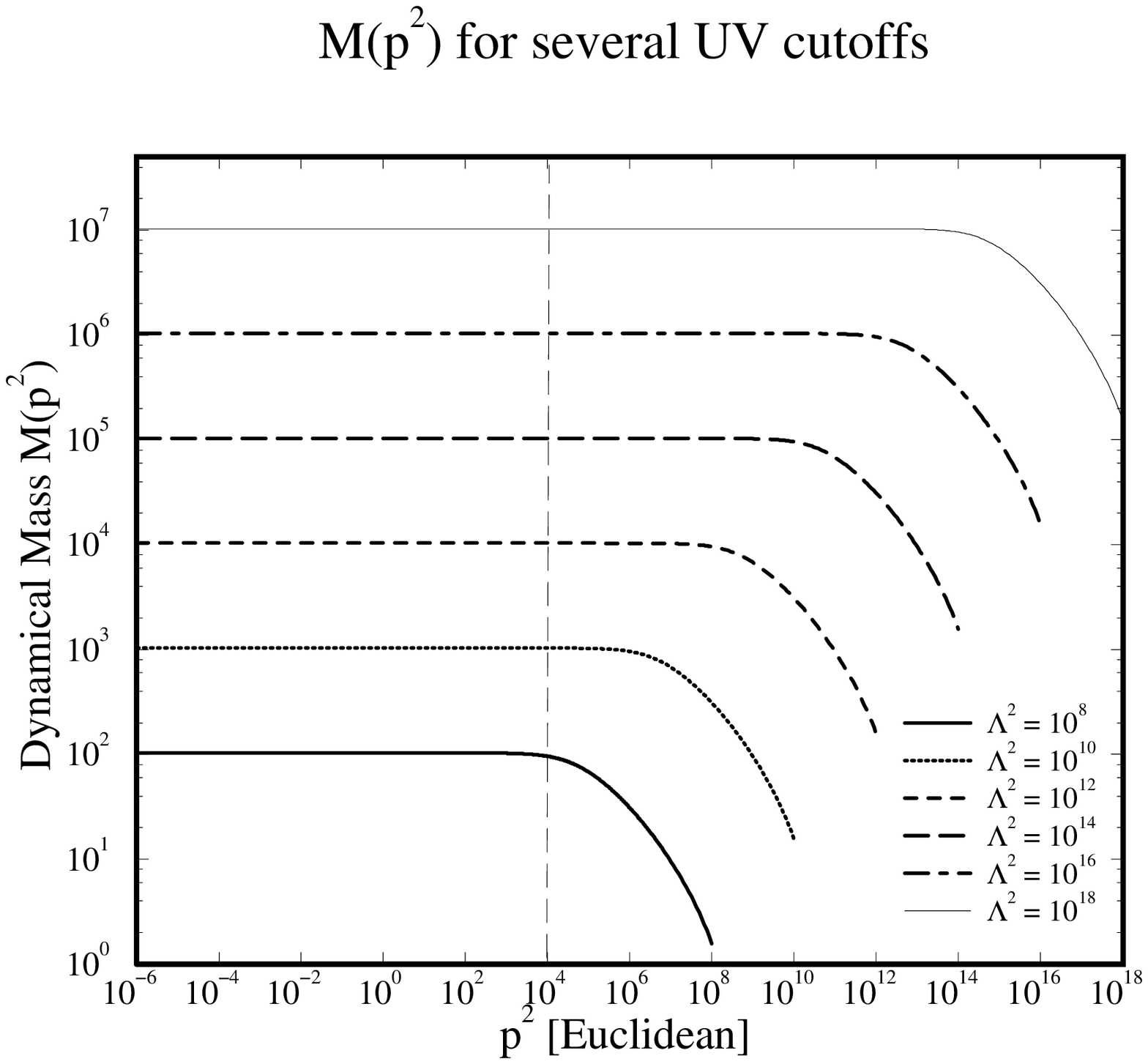}
  \caption{
        The behaviour of the finite renormalization $A(p^2)$ and the
        mass function $M(p^2)$ as a function of the ultraviolet
        cut-off $\Lambda$ for $m_0(\Lambda)=0$.  These solutions
        were for renormalization point $\mu^2=10^4$,
	coupling $\alpha=1.15$, and gauge parameter $\xi=0.25$.
	As $\Lambda\to\infty$ we find $A(p^2)\to 1$
	for all $p^2$ and $M(p^2)$ diverges with $\Lambda$.
  \label{Lambda_limit}}
\end{figure}

Fig.~\ref{Lambda_limit}
shows the non-oscillating solution $A(p^2)$ and $M(p^2)$
for supercritical coupling and with $m_0(\Lambda)=0$,
for a wide range of values of $\Lambda$.
It is clear from these numerical solutions that in the continuum limit
(i.e., $\Lambda\to\infty$) and for supercritical coupling,
we find $A(p^2)\to 1$ for all $p^2$ and a mass function $M(p^2)$
which diverges proportionally to $\Lambda$.
Conversely, for subcritical coupling (i.e., $\alpha<\alpha_c$),
the chiral limit does exist and we find simply that
  $A(\mu; p^2) = (p^2/\mu^2)^{-\alpha \xi / 4 \pi}$.
It seems clear from these numerical studies
that above critical coupling there is no finite chiral limit
in the continuum quenched theory in any covariant gauge,
even in the presence of the renormalization procedure.

While these conclusions were based on a numerical study with a
specific choice of vertex, one can easily construct
a general argument which applies irrespective of this choice:
  Consider any vertex $\Gamma$ which satisfies the WTI and
  which leads to multiplicative renormalizability \cite{qed4_hwr}.
  It automatically follows that $M(p^2)$ and $A(p^2)/Z_2(\mu,\Lambda)$
  are renormalization point independent, as in Eq.\ (\ref{ren_pt_transf}).
  We can then define dimensionless quantities by appropriately scaling
  with $\Lambda$, i.e., $\widehat\mu\equiv\mu/\Lambda$,
  $\widehat{p}^2\equiv p^2/\Lambda^2$, 
  $\widehat M(\widehat{p}^2)\equiv M(p^2)/\Lambda$,
  $\widehat A(\widehat{p}^2)\equiv A(\mu;p^2)/Z_2(\mu,\Lambda)$,
  and $\widehat{m}_0\equiv m_0(\Lambda)/\Lambda$.
  Note that the renormalization condition $A(\mu;\mu^2)=1$
  automatically determines $Z_2(\mu,\Lambda)$ for a given solution
  for fixed $\Lambda$.
  The dimensionless functions
  $\widehat{M}(\widehat{p}^2)$ and $\widehat{A}(\widehat{p}^2)$
  can only depend on dimensionless parameters, i.e., $\alpha$, $\xi$,
  $\widehat{\mu}$, and $\widehat{m}_0$. Furthermore, since for any
  fixed $\Lambda$ they are independent of $\mu$ (recall that we are
  working only in the quenched approximation), then it follows that
  they must in turn be independent of $\widehat{\mu}$.  Now for any
  finite $\Lambda$ and the choice $m_0(\Lambda)=0$ we have
  $\widehat{m}_0=0$.  Hence solving for any $\mu$ and $\Lambda$ with
  $m_0(\Lambda)=0$ allows us to form $\widehat{M}(\widehat{p}^2)$ and
  $\widehat{A}(\widehat{p}^2)$, from which we can read off the
  solutions for $A(p^2)$ and $M(p^2)$ for any other $\mu$ and $\Lambda$
  with vanishing bare mass using the above rules.
  We see then that the resulting $M(p^2)$ must diverge with $\Lambda$.
  So if $M(p^2) \not\equiv 0$, it must diverge with $\Lambda$
  as was found numerically.
  To summarize, above critical coupling
  any vertex which satisfies the WTI and leads to multiplicative
  renormalizability will lead to a diverging mass function
  in the chiral limit for quenched QED in four dimensions.  

\begin{figure}[thb]
  \hskip 2.5cm
  \setlength{\epsfxsize}{6.5cm}
      \epsffile{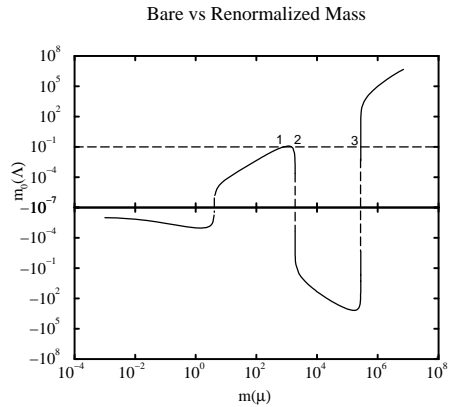}
  \caption{
        The relationship between the bare mass ($m_0(\Lambda)$)
	and the renormalized mass ($m(\mu)$) for the renormalized
	finite solutions.
        These result from solving for a given $m(\mu)$
	and extracting $m_0(\Lambda)$.
	The parameters for these solutions were the
	renormalization point $\mu^2=10^4$, $\alpha=1.25$,
	gauge parameter $\xi=0.25$ and $\Lambda^2=10^{14}$.
        The dashed horizontal line shows, e.g., that
	for $m_0(\Lambda)=0.1$ there are three solutions.  
        (The dashed vertical lines are merely to guide the eye
	on this back-to-back log scale.)
  \label{mmu_vs_m0}}
\end{figure}
\begin{figure}[htb]
  \hskip 2.5cm
  \setlength{\epsfxsize}{6.5cm}
      \epsffile{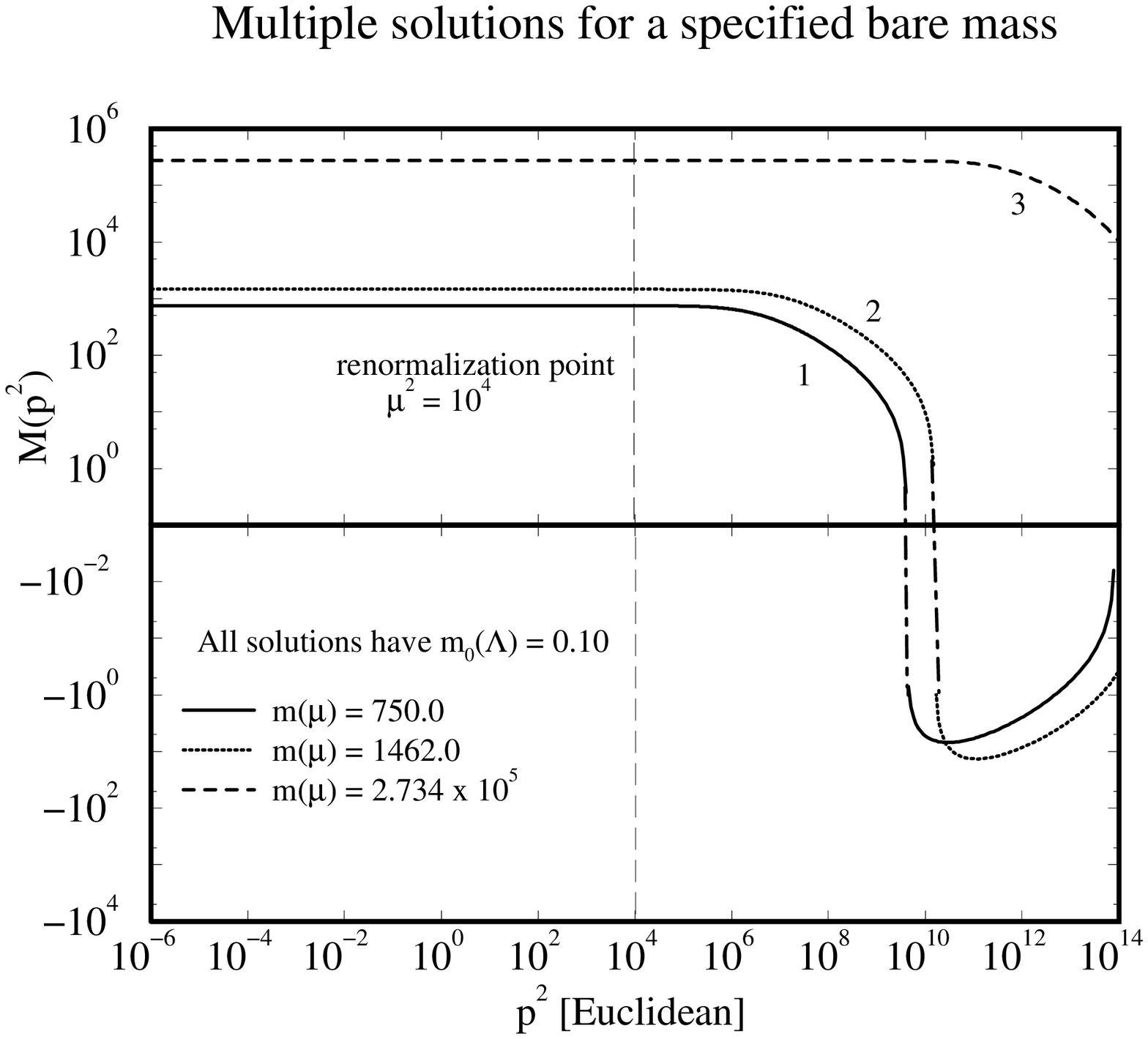}
  \caption{
        The three renormalized finite solutions
	corresponding to the same bare mass,
	i.e., $m_0(\Lambda)=0.1$, as indicated in Fig.~2.  
        The other parameters for these solutions were
	renormalization point $\mu^2=10^4$, coupling $\alpha=1.25$,
	gauge parameter $\xi=0.25$ and $\Lambda^2=10^{14}$.
        (The short-long dashed vertical lines
	connecting the upper and lower curves
	are merely to guide the eye on this back-to-back log scale.)
  \label{multiple_solutions}}
\end{figure}

It remains to discuss the meaning of the well-behaved finite solutions.
Our conventional thinking about the chiral limit would imply that
since ECSB should only increase the mass function above that found
in its absence,
then any solution above critical coupling which also has ECSB
must also correspond to a divergent mass in the continuum limit.
Thus the finite solutions do not correspond to the chiral limit
nor to any conventional concept of ECSB.
Further, the finite solutions show damped oscillations
in the dynamical mass function, periodic in $\ln{(p^2)}$,
and a corresponding oscillatory behavior for the bare mass.
Thus it is possible for a given set of parameters
  $\alpha$, $m_0(\Lambda)$, $\mu$, $\xi$, and $\Lambda$
to admit more than one finite solution.
We explicitly show this
in Figs.~\ref{mmu_vs_m0} and \ref{multiple_solutions},
where for a given set
  $\alpha$, $m_0(\Lambda)$, $\mu$, $\xi$, and $\Lambda$
there are distinct solutions.
As $\Lambda\to\infty$ the number of simultaneous solutions
becomes infinite.
  It seems likely that this behavior for the finite solutions
  is independent of the detailed vertex choice and furthermore
  that the same behavior can be induced in QCD
  by forcing the renormalized mass into an unphysical regime
  (by choosing $m(\mu)\equiv M(\mu^2)$ below the value
  corresponding to the chiral limit).
  It also seems likely that unquenching the theory will not remove
  this rather undesirable behaviour in QED$_4$,
  since the running coupling increases with scale
  rather than decreasing as in QCD.
  These latter conjectures are the subject of current investigation
  \cite{qcd_hsw}.


\section*{Acknowledgements}

This work was partially supported by the Australian Research Council,
by the U.S. Department of Energy
through Contract No.\ DE-FG05-86ER40273,
and by the Florida State University Supercomputer Computations Research
Institute which is partially funded by the Department of Energy
through Contract No.\ DE-FC05-85ER250000.
This research was also partly supported by grants of supercomputer time
from the U.S. National Energy Research Supercomputer Center
and the Australian National University Supercomputer Facility.


\end{document}